\documentclass[10pt,a4paper,english]{article}
\usepackage[english]{babel}
\usepackage[utf8]{inputenc}
\usepackage{amsmath}
\usepackage{amsfonts}
\usepackage{amssymb}
\usepackage{graphicx}
\usepackage{slashbox}

\newcommand{\be}{\begin{equation}}
\newcommand{\ee}{\end{equation}}

\newcommand{\aaa}{\alpha}
\newcommand{\piinsc}{\pi_0^\alpha}
\newcommand{\taa}{\tau^\alpha}

\newcommand{\saa}{\sigma^\alpha_0}

\begin{document}
\title{Results of three quantitative predictions based on past regularities about voter turnout at the French 2009 European election}
\author{Christian Borghesi\thanks{Electronic address: christian.borghesi@cea.fr}\\
\textit{Service de Physique de l'\'Etat Condens\'e, Orme des Merisiers,}\\ \textit{CEA Saclay, Gif sur Yvette Cedex, 91191, France.}}
\date {(June 15, 2009)}
\maketitle
\begin{abstract}
\normalsize
Twelve turnout rates of French national elections by municipality have shown statistical regularities, neither depend on the nature of the election, nor on the national turnout level. Three quantitative predictions about voter turnout at the French 2009 European election were made in arXiv:physics/0905.4578. Here, we give the results of these three predictions. Each one is confirmed by real measures.
\end{abstract}

A physicist overview of electoral studies is made in~\cite{fortunato_stat_phys}. In this paper we give the results of the three predictions made in~\cite{three_predic} about the voter turnout of the French 2009 European Parliament election. 

This election was held on June 7, 2009. For metropolitan France, there were $42.4~10^6$ registered voters~\cite{bureau}, and the turnout rate was $41.4\%$. Notations in this paper are the same as in~\cite{three_predic}.\\
\\
\large{\textbf{1. Standard deviation of $\taa$}}\\

\begin{table}[h!]
\hspace{0.5cm}Previous, predicted, and real standard deviation of $\taa$ (see section 1 in~\cite{three_predic}) over all the $\aaa$ municipalities, are:
\begin{center}
\begin{tabular}{|c|c|c|c|}
\cline{2-4}
\multicolumn{1}{c|}{}
 & Previous measures & Expected measure & Real value\\
\hline
Standard deviation of $\taa$ & $0.376 \pm 0.019$ & $[0.338 ; 0.414]$ & 0.360\\
\hline
\end{tabular}\\
\begin{flushleft}(In the above table like in the next two ones, \textit{Previous measures} is written as $(mean\pm~standard\;deviation)$, and the prediction is given within an arbitrary \textit{two sigma}.)\end{flushleft}
\end{center}
\end{table}

\large{\textbf{2. {Correlation of $\mathbf{\saa}$ at different elections}}\\

\begin{table}[h!]
\hspace{0.5cm}Previous $C_{t_i,t_j}(\sigma_0)$ for all couples of different elections (see section 2 in~\cite{three_predic}); predicted and real averages of $C_{t_i,t_j}(\sigma_0)$ for couples constituted by the 2009 European election and one of the twelve former elections, are:
\begin{center}
\begin{tabular}{|c|c|c|c|}
\cline{2-4}
\multicolumn{1}{c|}{}
& Previous measures & Expected measure & Real measure\\
\hline
$C_{t_i,t_j}(\sigma_0)$ & $0.567 \pm 0.058$ & $[0.451 ; 0.683]$ & $0.577$\\
\hline
\end{tabular}
\begin{flushleft}(Tab.~\ref{ttempo-saa-abst-09} gives the twelve $C_{t_i,t_j}(\sigma_0)$, where $t_i$ is one of the past twelve elections and, $t_j$, the 2009 European Parliament election.)\end{flushleft}
\end{center}
\end{table}

\large{\textbf{3. {Correlation between $\mathbf{\saa}$ and $\mathbf{\piinsc}$}}\\

\begin{table}[h!]
\hspace{0.5cm}Previous, predicted and real correlation between $\saa$ and $\piinsc$ (see section 3 in~\cite{three_predic}) are:
\begin{center}
\begin{tabular}{|c|c|c|c|}
\cline{2-4}
\multicolumn{1}{c|}{}
 & Previous measures & Expected measure & Real measure\\
\hline
Correlation between $\saa$ and $\piinsc$ & $0.645 \pm 0.026$ & $[0.593 ; 0.697]$ & 0.657\\
\hline
\end{tabular}
\end{center}
\end{table}

\vspace{0.5cm}
\cite{three_predic} dealt with a relatively simple phenomenon: the participation in French national elections. Three regularities have been observed for the past twelve turnout rates per municipality (or ``commune'' of France). Three predictions based on past statistical regularities were made for the French 2009 European Parliament election. Real results are in agreement with these predictions.

\vspace{0.5cm}
{\bf Acknowledgments\\}
I would like to thank Brigitte Hazart who has sent me electoral data as soon as possible, and also Lionel Tabourier for his help.

\vspace{0.5cm}
\begin{table}[h!]
\begin{tabular}{|c|c|c|c|c|c|c|c|c|c|c|c|}
\hline
92-b & 94-m & 95-m & 95-b & 99-m & 00-b & 02-m & 02-b & 04-m & 05-b & 07-m & 07-b\\
\hline
0.534 &  0.585 & 0.533 &  0.529 & 0.648 &  0.622 &  0.539 & 0.537 & 0.679 &  0.596 &  0.532 &  0.589\\
\hline
\end{tabular}
\caption{\small $C_{t_i,t_j}(\sigma_0)$, where $t_i$ is one of the previous twelve elections and $t_j$, the 2009 European Parliament election. The mean value is $0.577$.}
\label{ttempo-saa-abst-09}
\end{table}

\end{document}